\documentclass[conference]{IEEEtran}
\usepackage[T1]{fontenc}
\usepackage[utf8]{inputenc}
\usepackage{amsmath,amssymb,amsthm}
\usepackage{graphicx}
\usepackage{cite}

\title{Signal Constellations with Enhanced Energy Efficiency for High-Speed Communication Systems}

\author{
\IEEEauthorblockN{Mark Bykhovskiy, Member, IEEE}
\IEEEauthorblockA{
Moscow Technical University of Communications and Informatics (MTUCI), Moscow, Russia\\
e-mail: bykhmark@gmail.com
}
}

\newtheorem{theorem}{Theorem}

\begin{document}
\maketitle

\begin{abstract}
This paper proposes a new method for constructing multidimensional signal constellations (SC), referred to as SCOPT, for high-speed communication systems with enhanced energy efficiency (EE).

In contrast to conventional approaches, the proposed method increases the minimum Euclidean distance (MED) between signals by increasing the normalized signal duration, without relying on coding or increasing transmit power.

Analytical expressions for the demodulation error probability and the energy loss relative to the Shannon limit are derived. It is shown that, unlike classical Shannon-type constellations (SCSH), SCOPT enable reliable communication regimes in which the required signal-to-noise ratio may fall below the conventional Shannon limit within the adopted geometric framework.

The proposed constellations retain a simple structure compatible with standard modulation schemes such as QAM and APSK, making them suitable for practical implementation in modern communication systems.

Numerical analysis demonstrates that SCOPT significantly outperform SCSH in terms of energy efficiency while requiring substantially shorter signal duration.
\end{abstract}


\section{Introduction}

The development of telecommunication systems has been, and continues to be, one of the key factors in the progress of human civilization. This section considers the fundamental concepts of modern communication theory underlying the transmission of information over communication channels. These concepts are based on the ideas of Claude Shannon, the founder of information theory, presented in his seminal work \cite{ref1}.

Shannon’s works \cite{ref2,ref3}, devoted to the construction of signal constellations providing high energy efficiency under a given reliability constraint, serve as the starting point of this study.

A message is represented as a sequence of $N$ digital symbols, each of which can take one of $q$ possible values. For their transmission over a communication channel, $N$-dimensional signals are used. These signals are typically harmonic and consist of $n$ orthogonal frequencies. Each frequency has two orthogonal components (sine and cosine), so that $N = 2n$.

In \cite{ref2}, Shannon showed that each message can be associated with a signal point in an $N$-dimensional Euclidean signal space, where $N = 2FT$, $F$ is the channel bandwidth, $T$ is the signal duration, and $n = FT$ is the normalized signal duration. A similar geometric representation of signals was previously used by V. A. Kotelnikov \cite{ref4}.

The geometric interpretation of signals in multidimensional space is one of the key tools of communication theory. Each message corresponds to a vector whose coordinates are determined by the parameters of the $N$-dimensional signal. To transmit $M$ different messages, a signal constellation (SC) consisting of $M$ distinct signals must be constructed.

When receiving a signal, the presence of Gaussian noise leads to a displacement of the signal point. The reliability of reception is determined by the demodulation error probability $P_{\mathrm{dem}}$, which typically lies in the range $10^{-7}$--$10^{-12}$.

In constructing SC, it is necessary to maximize the minimum Euclidean distance (MED) between signals, defined as
\begin{equation}
D_E(n,\rho_s)=
\sqrt{
\min_{m_1,m_2}
\left\{
\frac{1}{T}\int_0^T [S_{m_1}(t)-S_{m_2}(t)]^2\,dt
\right\}
}.
\tag{1}
\end{equation}

The value of MED depends on the normalized signal duration $n$, the signal-to-noise ratio $\rho_s$, and, most importantly, on the geometric structure of the constellation, in particular on the shape of the decision region (DR).

Shannon was the first to investigate signal constellations that improve the energy efficiency of communication systems. Such constellations are denoted as SCSH.

The key feature of SCSH is that the increase of MED is achieved not by increasing transmitter power, but by increasing the normalized signal duration $n$. Another important feature is that the decision region has the form of an $N$-dimensional sphere. A third feature is that signal coordinates are determined only by information symbols, and error-correcting codes (ECC) are not used.

In \cite{ref2}, Shannon obtained the fundamental relation
\begin{equation}
R_f=\log_2(1+\rho_{s0}),
\tag{2}
\end{equation}
where $\rho_{s0}$ is the Shannon limit.

According to Shannon’s theorem, reliable communication is possible only if $\rho_s > \rho_{s0}$; otherwise, reliable transmission is impossible.

In systems with spectral efficiency $R_f$, the energy efficiency is evaluated by $\Delta\rho_s(n)=10\log_{10}(\rho_s/\rho_{s0})>0$ dB.

In this paper, an alternative class of signal constellations, denoted as SCOPT, is investigated.

The principal difference of the proposed approach from existing methods is that the increase of the minimum Euclidean distance is achieved not by increasing transmitter power or by using coding, but by increasing the normalized signal duration under a different geometric structure of the signal constellation. This makes it possible to realize operating regimes in which the required signal-to-noise ratio may be lower than the classical Shannon limit, which was previously considered impossible within conventional models.

The remainder of the paper is organized as follows. Section II provides a brief overview of existing methods for constructing multidimensional signal constellations, including a comparison of their energy losses (EL) relative to SCSH, as well as an analysis of the required signal duration for achieving a given reliability. Section III investigates the properties of SCSH and derives expressions for $P_{\mathrm{dem}}$ and $\Delta\rho_s(n)$. Section IV presents the construction method of SCOPT and derives the corresponding analytical expressions. Section V compares the energy efficiency of communication systems employing SCSH and SCOPT. Section VI concludes the paper and outlines directions for future research.

\section{Methods for Constructing Signal Constellations}

A significant number of studies have been devoted to the construction of signal constellations (SC) for high-speed communication systems whose energy efficiency (EE) and spectral efficiency (SE) are close to those of SCSH.

Among the most important are works \cite{ref5,ref6,ref7}, in which the problem of dense packing of signal points in multidimensional space was investigated. In these studies, one- and two-dimensional constellations such as pulse-amplitude modulation (PAM), quadrature amplitude modulation (QAM), amplitude phase shift modulation (APSK), and hexagonal modulation (HEXM) were used as basic elements.

The use of QAM and APSK in modern high-speed communication systems is standardized, for example, in digital broadcasting systems DVB-T2 and DVB-S2.

It should be noted that Shannon did not propose a constructive method for building signal constellations of the SCSH type. The use of the Hartley approach for estimating the number of signals led to the idea of constructing such constellations based on packing multidimensional spheres in signal space.

During the period 1980--2000, numerous studies were carried out in this direction (see, for example, \cite{ref8}). However, analysis of \cite{ref6,ref8} shows that the dimensionality of such constellations typically does not exceed $N=24$, and their spectral efficiency does not exceed $R_f \approx 2$ bit/s$\cdot$Hz. With such a limited dimensionality, it is impossible to approach the performance of Shannon’s ``ideal'' communication system.

In \cite{ref6,ref8}, signal constellations based on coded modulation were also investigated.

In modern coding schemes, each codeword (CW) consists of $N$ symbols, of which $k$ are information symbols and $r$ are redundancy (parity-check) symbols, so that $N = k + r$. Each symbol can take one of $q$ possible values; therefore, the number of possible transmitted signals is $M = q^k$. The key parameters of such codes are the code rate $R_c = k/N < 1$ and the minimum Hamming distance $D_c(N, R_c)$, defined as the number of differing symbols between codewords.

Codewords are transmitted using one of the basic signal constellations (e.g., QAM or APSK). In this case, each CW can be considered as one signal of a constellation denoted as SCCOD. For SCCOD, the minimum Euclidean distance between signals is given by $D_{EC}(\rho_s,R_f,R_c)=D_E(\rho_s,R_f)\,D_c(N,R_c)$.

The construction of codes and coded modulation schemes is described in detail in \cite{ref5,ref6,ref7,ref9,ref10}. Various classes of codes have been developed for SCCOD, including Reed--Solomon codes, low-density parity-check (LDPC) codes, turbo codes, convolutional codes, and polar codes.

Based on these codes, different coded modulation techniques have been proposed, such as trellis coded modulation (TCM), multilevel coded modulation (MCM), and bit-interleaved coded modulation (BICM).

It is well known that without coding, the improvement in EE of traditional constellations (QAM, APSK, HEXM) is limited to approximately 1 dB. The use of coding and coded modulation allows a significantly larger gain.

The spectral efficiency of coded constellations is given by $R_{fc}=R_fR_c<R_f$.

In \cite{ref11}, the energy loss (EL) of SCCOD relative to the Shannon limit was analyzed for systems employing Reed--Solomon codes. It was shown that the optimal operating regime corresponds to a demodulation error probability of approximately $p_{er}(\rho_s)\approx0.5\cdot10^{-2}$--$10^{-1}$. In this case, it is reasonable to use codes with high rate Rc ~ 0.7-0.9. Under these conditions, the energy losses are approximately: 2-3 dB for Rfc = 1-3 bit/s$\cdot$Hz and 5-6 dB for Rfc = 5-10 bit/s$\cdot$Hz. For higher spectral efficiency (above 10 bit/s$\cdot$Hz), the energy losses may exceed 10 dB.

A major drawback of SCCOD is that, in order to ensure high reliability of message reception, the number of symbols in a codeword must be very large (on the order of 10,000--60,000 symbols). This leads to significant transmission delay and increased complexity of the decoding device.

At present, the use of SCCOD remains the primary method for ensuring reliable communication and improving EE in practical systems.

However, analysis of modern communication systems using SCCOD and systems based on SCSH, presented in \cite{ref11}, shows that SCCOD not only exhibit significant energy losses relative to SCSH, but also require substantially larger normalized signal durations---often up to 30 times greater.

As a result, the performance of modern communication systems remains far from Shannon’s ``ideal'' systems. In our opinion, this is due to the fact that Shannon’s key idea---improving EE by increasing the normalized signal duration---has not been effectively utilized in the construction of signal constellations.

\section{Probability of Error for SCSH}

In [2], C. Shannon proposed a geometric approach to the construction of signal constellations (SC), which makes it possible to realize high-speed communication systems with high reception reliability. In the same work, a method for estimating the error probability was introduced, known as the random coding method. This method makes it possible to obtain an upper bound on the average error probability over an ensemble of randomly selected SC. This method is widely used for estimating the demodulation error probability $P_{\mathrm{dem}}$ in cases where a constructive algorithm for building SC is unknown, but its main parameters---namely, the number of signals and the minimum distance between them---are specified. A detailed description of this method is given, for example, in [13].

In the geometric model of SCSH [2], it is assumed that each signal is surrounded by a spherical decision region (DR) of radius $\sqrt{n}$. The nominal points of the $N$-dimensional signals are located on the surface of a sphere of radius $\sqrt{n\rho_s}$. Under the action of additive Gaussian noise, the received signal is randomly displaced from its nominal position to the surface of a sphere of radius $\sqrt{n(1+\rho_s)}$.

The number of signals providing spectral efficiency $R_f$ is determined by the expression $M=(1+\rho_{s0})^n$, where $\rho_{s0}=(2^{R_f}-1)$ is the Shannon limit. The minimum Euclidean distance (MED) between signals in SCSH is determined as follows [2]
\begin{equation}
\sqrt{D_E(\rho_s,\rho_{s0},n)}=2\sqrt{n(\rho_s/\rho_{s0})}.
\tag{3}
\end{equation}

A reception error occurs when the noise displacement causes the signal to leave its corresponding decision region. Based on this model, Shannon obtained the following estimate of the error probability:
\begin{equation}
P_{\mathrm{dem}}(n,\rho_s,R_f)=
\begin{cases}
e^{-n\ln\left(\frac{1+\rho_s}{1+\rho_{s0}}\right)}, & (\rho_s/\rho_{s0})>1,\\
1, & (\rho_s/\rho_{s0})\le 1.
\end{cases}
\tag{4}
\end{equation}

Using (4), the energy loss (EL) relative to the Shannon limit can be determined as
\begin{equation}
\Delta\rho_s(n)=10\log_{10}\left\{\frac{\left[2^{R_f}/\sqrt[n]{P_{\mathrm{dem}}}\right]-1}{2^{R_f}-1}\right\}.
\tag{5}
\end{equation}

Expression (4) directly yields Shannon's theorem and the corresponding expression for the limiting spectral efficiency. The random coding method was also used in [3], [13]--[15] to obtain other estimates of $P_{\mathrm{dem}}$.

In this section, we derive an alternative expression for the error probability using a more illustrative approach, similar to classical methods in communication theory, as described, for example, in [9]. This method of proving Shannon's theorem is significantly simpler than those based on the random coding method.

Let us assume that the decision region has the form of an $N$-dimensional sphere. Under the action of noise, the signal position is randomly displaced, and the magnitude of the displacement is $Z_n=\sqrt{\sum_{i=1}^{N} n_i^2}$, where $n_i$ are independent Gaussian random variables with zero mean and variance 0.5; the square of $Z_n$ has the distribution $p(z)=\frac{z^{n-1}e^{-z}}{(n-1)!}$, and its mean value is equal to $n$, and as $n\to\infty$, the distribution becomes concentrated near its mean value, which makes it possible to interpret the sphere of radius $\sqrt{n}$ as a region of uncertainty.

The condition for the occurrence of an error is
\begin{equation}
z=\left(\sum_{i=1}^{N} n_i^2\right)\ge D_E(\rho_s,\rho_{s0},n)/4=n(\rho_s/\rho_{s0}).
\tag{6}
\end{equation}

The error probability is determined by the integral $P_{\mathrm{dem}}(\rho_s,R_f,n)=\int_{n(\rho_s/\rho_{s0})}^{\infty} p_n(z)\,dz$.

Using the Chernoff method [9], we obtain the estimate
\begin{equation}
P_{\mathrm{dem}}(\rho_s,R_f,n)\cong
\begin{cases}
\exp\{-n[y(\rho_s/\rho_{s0})]\}, & (\rho_s/\rho_{s0})>1,\\
1, & (\rho_s/\rho_{s0})\le 1,
\end{cases}
\tag{7}
\end{equation}
where $y(\rho_s/\rho_{s0})=(\rho_s/\rho_{s0})-1-\ln(\rho_s/\rho_{s0})$.

This expression gives a sufficiently accurate upper bound for the error probability. From (7) it follows that when $(\rho_s/\rho_{s0})>1$, one has $P_{\mathrm{dem}}(\rho_s,\rho_{s0},n)\to 0$ as $n\to\infty$, which is consistent with Shannon's theorem. The dependencies (4) and (7) exhibit a pronounced threshold behavior: when $(\rho_s/\rho_{s0})\le 1$, the error probability is close to unity, whereas for $(\rho_s/\rho_{s0})>1$ it rapidly decreases with increasing $n$.

For practical calculations, it is convenient to use the dependence of the energy loss on $P_{\mathrm{dem}}$ and $n$. This dependence is obtained via the inverse function of $y(u)=u-1-\ln(u)$. For the range $1\le u\le 10$, the following approximation is used with high accuracy: $u(y)=1+\frac{y}{2}+\sqrt{3y+\left(\frac{y}{2}\right)^2}$.

Then the energy loss is given by
\begin{equation}
\Delta\rho_s(n)=10\log_{10}\left[u\left(\frac{\ln(1/P_{\mathrm{dem}})}{n}\right)\right].
\tag{8}
\end{equation}

From (8) it follows that for finite values of $n$, systems with SCSH inevitably exhibit energy losses relative to the ``ideal'' Shannon system. As $n\to\infty$, $\Delta\rho_s(n)\to 0$ dB. Analysis shows that expressions (5) and (8) yield close results (with a discrepancy of about 1 dB).

An important feature of SCSH is that the condition of correct reception is determined by the total noise energy across all signal coordinates. This means that the decision must be made for the entire $N$-dimensional vector, which significantly complicates the demodulation procedure. In addition, from condition (6) it follows that when $(\rho_s/\rho_{s0})\le 1$, reliable reception is impossible, which limits the energy efficiency of SCSH constellations by the Shannon limit.

In [3], Shannon considered SCSH constellations in which signal points are uniformly distributed over the surface of an $N$-dimensional sphere of radius $\sqrt{n\rho_s}$, while the decision region has the form of an $N$-dimensional cone. Using the random coding method, estimates of the error probability were obtained in [3], but their derivation is based on rather complex analysis. In [12], analogous estimates were obtained using the same simple method as that used in deriving (7). These expressions have almost the same form as (7) and yield very close numerical results. This is natural, since for $n\gg 1$, almost the entire volume of the signal space for SCSH is concentrated near the surface of the corresponding hypersphere.

\section{Probability of Error for SCOPT}

In Shannon's works [2], [3], the structure of SCSH-type constellations was not specified constructively; only their main parameters were defined. In this section, the structure of one possible SCOPT constellation is described, whose elements are one-dimensional PAM signals.

The signals belonging to the SCOPT constellation and applied to the input of the receiver demodulator can be represented as
\begin{equation}
S_m(t)=\sqrt{n\rho_s}\left[\sum_{l=1}^{q} S_{ml}(t)\right].
\tag{9}
\end{equation}

Here $S_{ml}(t)=\left[\sum_{i=1}^{2K} x_{(l-1)q+i}^{m} V_{(l-1)q+i}(t)\right]$, where $V_{(l-1)q+i}(t)$ are elementary orthogonal signals used for transmitting information symbols $x_{(l-1)q+i}^{m}$ [9]. Such signals may be harmonic functions or other orthogonal signals, for example, pseudorandom sequences [9].

The signals $\hat{S}_m(t)=S_m(t)/\sqrt{n\rho_s}$ and $S_{ml}(t)$ are normalized, so that their average power is equal to unity. From (9) it follows that the $N$-dimensional signal $S_m(t)$ is a sum of $q$ signals $S_{ml}(t)$, each having dimensionality $2K$. Therefore, $N=2Kq$, and the normalized signal duration is $n=Kq$.

The symbols $x_{(l-1)q+i}^{m}$ define the normalized Euclidean coordinates of the signal points in the $N$-dimensional space. Unlike SCSH, the SCOPT signals are not required to lie inside an $N$-dimensional sphere. Each signal $S_{ml}(t)$ contains $2K$ coordinates whose values differ by $d_E$, which represents the normalized Euclidean distance between adjacent levels. The coordinates take values $x_{(l-1)q+i}^{m}(l)=0.5d_E[2(l-K)-1]$, $l=1,2,\dots,2K$, i.e., they lie in the interval $x_{(l-1)q+i}^{m}(l)\in[-0.5d_E(2K-1),\,0.5d_E(2K-1)]$. Different signals $S_{ml}(t)$ differ by permutations of coordinates with different values. The number of such permutations is $M_{sl}=(2K)!$; therefore the total number of signals in the SCOPT constellation is $M_s=(M_{sl})^q$.

Accordingly, the spectral efficiency is
\begin{equation}
R_f(K)=\log_2(M_s)/Kq=\log_2[(2K)!]/K\cong \log_2(4K^2).
\tag{10}
\end{equation}

The approximation follows from Stirling's formula.

To determine the average power of the normalized signal $\hat{S}_m(t)$, we use the expression $P_m=\int_0^T [\hat{S}_m(t)]^2 dt=(d_E^2/4)\,[q\sum_{i=1}^{2K}(2i-1)^2]/qK$. Hence $P_m=(d_E^2/2)[(4K^2-1)/3]$. From the condition $P_m=1$, we obtain $d_E=\sqrt{6/(4K^2-1)}$.

Thus, the minimum Euclidean distance between signals in the SCOPT constellation is
\begin{equation}
D_E(\rho_s,R_f,n)=\sqrt{6n[\rho_s/(4K^2-1)]}.
\tag{11}
\end{equation}

The fundamental difference between SCSH and SCOPT lies in the way noise affects the signal position in the $N$-dimensional space. For SCSH, the condition for leaving the decision region is determined by the total effect of noise on all coordinates. In SCOPT constellations, noise affects each coordinate independently, and the coordinate values can be estimated independently in the demodulator. This significantly simplifies the demodulation procedure.

In this case, the decision region has the form of an $N$-dimensional hypercube, and the condition for an error in one coordinate is
\begin{equation}
|n_i|>D_E(\rho_s,R_f,n)/2.
\tag{12}
\end{equation}

The probability of error is equal to the probability that at least one coordinate leaves its decision region. Therefore, $P_{\mathrm{dem}}(\rho_s,R_f,n)=1-[1-2Q(D_E/2)]^{2n}$, where $Q(x)=\int_x^\infty \frac{e^{-0.5y^2}}{\sqrt{2\pi}}\,dy$.

For large $n$ and small $Q(D_E/2)$, this can be approximated as
\begin{equation}
P_{\mathrm{dem}}(\rho_s,R_f,n)\cong 4nQ(D_E/2).
\tag{13}
\end{equation}

From (13) it follows that for fixed $\rho_s$ and $R_f$, the error probability decreases with increasing signal duration, proportional to the parameter $q$.

\begin{equation}
\Delta\rho_s(n)=10\log_{10}\left\{\frac{(4K^2-1)\Phi(n,P_{\mathrm{dem}})}{6(2^{R_f(K)}-1)}\right\},
\tag{14}
\end{equation}
where $\Phi(n,P_{\mathrm{dem}})=\ln\!\left(\frac{Z(n,P_{\mathrm{dem}})}{\ln[Z_L(n,P_{\mathrm{dem}})]}\right)\!/n$, $Z(n,P_{\mathrm{dem}})=\left(n/(P_{\mathrm{dem}}\sqrt{2\pi})\right)^2$, and $Z_L(n,P_{\mathrm{dem}})=\ln[Z(n,P_{\mathrm{dem}})]$; the quantity $R_f(K)$ is defined by (10).

From (13) it follows that for SCOPT constellations, the energy loss coefficient $\Delta\rho_s(n)$ decreases with the parameter $n=qK$. However, unlike SCSH constellations, for which $\Delta\rho_s(n)\to 0$ dB as $n\to\infty$ (see (8)), in systems with SCOPT the value of $\Delta\rho_s(n)$ may become negative.

This means that communication systems using SCOPT can operate in regimes where the required signal-to-noise ratio is below the classical Shannon limit, while high reliability of message reception is ensured even for relatively small signal durations corresponding to $\Delta\rho_s(n)=0$ dB.

The main result of this section can be formulated as follows.

\begin{theorem}
In communication systems employing SCOPT, the error probability $P_{\mathrm{dem}}(\rho_s,R_f,n)$, which determines the reliability of message reception, decreases monotonically with increasing normalized signal duration $n$ for any value of the signal-to-noise ratio $\rho_s$ at the receiver input and for any spectral efficiency $R_f$. Therefore, reliable communication can be achieved even at arbitrarily low SNR, provided that the signal duration is sufficiently large.
\end{theorem}

In the next section, a quantitative comparison of the energy efficiency of systems using SCSH and SCOPT is presented.

\section{Comparison of Communication Systems Using SCSH and SCOPT}

To compare the performance of the signal constellations SCSH and SCOPT, Fig.~1 shows the dependences of the message reception error probability $P_{\mathrm{dem}}(\rho_s,R_f,n)$ on the signal-to-noise ratio $\rho_s$.

For SCSH constellations, the dependences $P_{\mathrm{dem}}(\rho_s,R_f,n)$ in Fig.~1 are plotted for $n=10^5$. For SCOPT, the results are shown for the following parameters: $R_f=2$ bit/s$\cdot$Hz, $n=qK=50$; $R_f=4$ bit/s$\cdot$Hz, $n=50$; $R_f=6$ bit/s$\cdot$Hz, $n=30$; and $R_f=8$ bit/s$\cdot$Hz, $n=20$.

\begin{figure}[!t]
\centering
\includegraphics[width=\columnwidth]{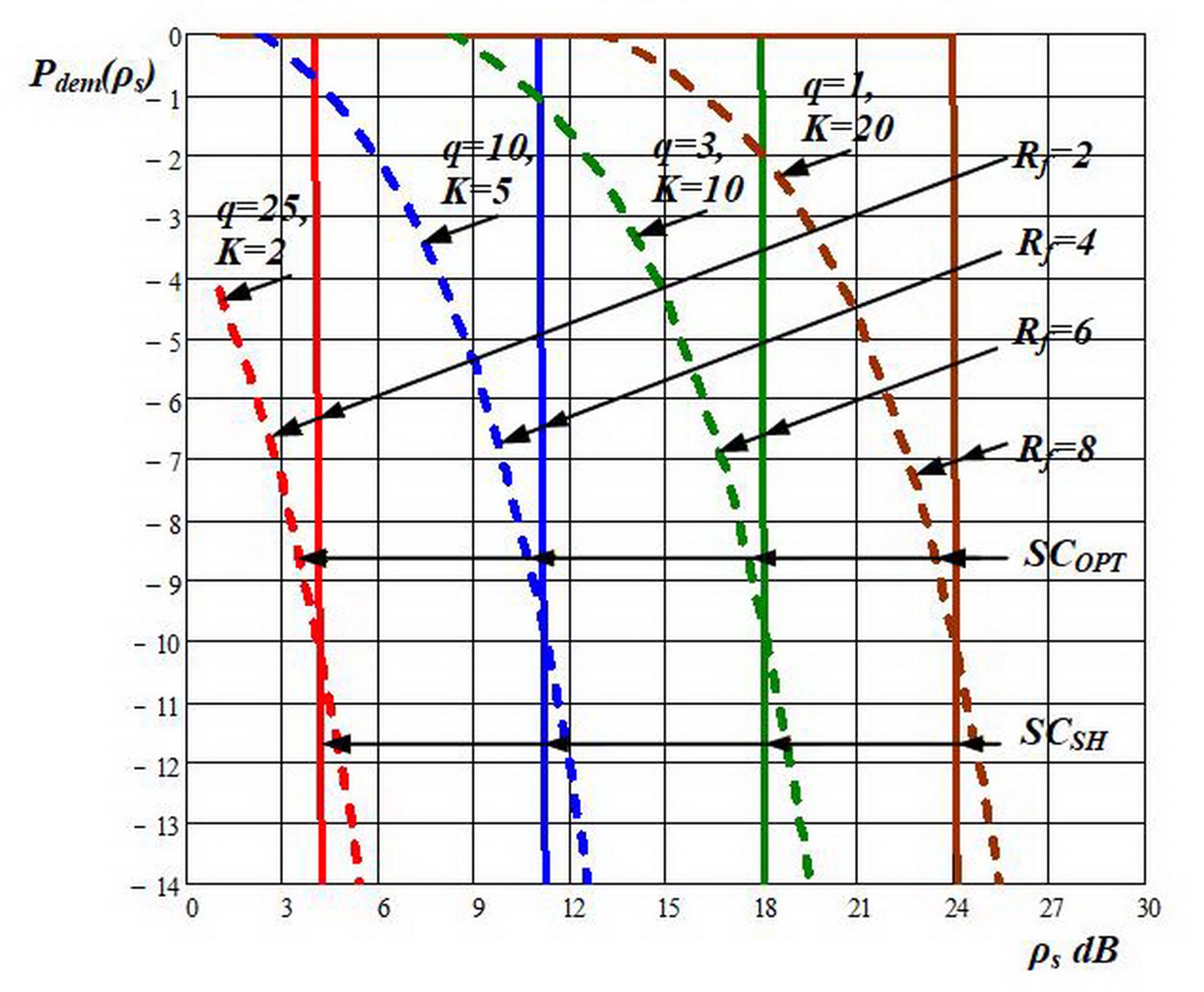}
\caption{Dependences of $P_{\mathrm{dem}}(\rho_s,R_f,n)$ on $\rho_s$ for SCSH and SCOPT.}
\end{figure}

The obtained results show that, for SCOPT, in contrast to SCSH, the dependence of the error probability $P_{\mathrm{dem}}(\rho_s,R_f,n)$ on the signal-to-noise ratio has a smoother behavior. In particular, even when $\rho_s<\rho_{s0}$, the error probability may take small values.

This indicates that the energy efficiency (EE) of communication systems using SCOPT is not limited by the classical Shannon limit derived for constellations of the SCSH type within the adopted geometric model.

It should also be noted that even for a normalized signal duration in SCSH much smaller than $n=10^5$, for example at $n=1000$, the energy loss (EL) relative to the Shannon limit is on the order of 1.4 dB. At the same time, the signal duration in systems with SCSH is about an order of magnitude larger than in systems with SCOPT.

These results confirm that the use of SCOPT makes it possible to significantly improve the EE of communication systems with much shorter signal duration. An additional important advantage of SCOPT is that increasing the signal duration by a factor of $q$ leads to a reduction of the required transmitter power by approximately $10\log_{10}(q)$ dB.

This has important practical significance, since it reduces the energy requirements of transmitting devices in modern communication systems.

\section{Conclusion}

The analysis of methods for constructing signal constellations with coded modulation (SCCOD), used in modern high-speed communication systems, has shown that, in order to ensure high reception reliability, such systems require the mandatory use of error-correcting codes (ECC) with long codewords. This substantially increases implementation complexity and leads to significant transmission delay.

The use of SCCOD is accompanied by a reduction in the energy efficiency (EE) of communication systems. The energy losses (EL) relative to the Shannon limit are typically 5 dB or more, and they increase with increasing spectral efficiency (SE).

This paper has proposed a method for constructing SCOPT signal constellations which, unlike SCSH constellations, have a simpler structure and are convenient for practical implementation. In terms of implementation complexity, communication systems using SCOPT are comparable to systems employing conventional QAM, APSK, or HEXM signals without coding. It has been shown that SCOPT provide substantially higher EE than SCSH.

The use of SCOPT appears promising in both terrestrial and satellite communication systems, including prospective 6G systems. The use of SCOPT makes it possible to reduce the required transmitter power by increasing the signal duration. At the same time, the signal duration required to ensure high communication reliability is significantly smaller than in systems using SCSH and SCCOD, which results in low transmission delay.

Promising directions for further research include the development of new signal constellations for both high-speed and low-speed communication systems ($R_f \ll 1$), as well as the design of constellations with reduced peak factor. Of particular interest is the development of SC for channels with multipath propagation and deep fading.

It should be emphasized that the geometric structure of signal constellations, in particular the shape of the decision region (DR), plays a key role in the construction of communication systems with high energy efficiency.

\end{document}